\documentclass[useAMS,usenatbib]{mn2e} 
\usepackage{graphicx}
\usepackage{amssymb} 
\usepackage{amsmath} 
\title{A possible radio signal of annihilating dark matter in the Abell 4038 cluster} 
\author[Chan \& Lee]{Man Ho Chan \thanks{chanmh@eduhk.hk}, Chak Man Lee
\\ Department of Science and Environmental Studies, The Education University of Hong Kong, Tai Po, Hong Kong}

\begin{document}

\date{Accepted XXXX, Received XXXX}

\pagerange{\pageref{firstpage}--\pageref{lastpage}} \pubyear{XXXX}

\maketitle

\label{firstpage}

\date{\today}

\begin{abstract}
In the past decade, some telescopes (e.g. Fermi-LAT, AMS and DAMPE) were launched to detect the signals of annihilating dark matter in our Galaxy. Although some excess of gamma rays, anti-protons and electrons/positrons have been reported and claimed as dark matter signals, the uncertainties of Galactic pulsars' contributions are still too large to confirm the claims. In this article, we report a possible radio signal of annihilating dark matter manifested in the archival radio continuum spectral data of the Abell 4038 cluster. By assuming the thermal annihilation cross section and comparing the dark matter annihilation model with the null hypothesis (cosmic ray emission without dark matter annihilation), we get very large Test Statistic values TS $>45$ for four popular annihilation channels, which correspond to more than $6 \sigma$ statistical preference. This reveals a possible potential signal of annihilating dark matter. In particular, our results are also consistent with the recent claims of dark matter mass $m \approx 30-50$ GeV annihilating via the $b\bar{b}$ quark channel with the thermal annihilation cross section.
\end{abstract}

\begin{keywords}
Dark matter
\end{keywords}

\section{Introduction}
The existence of dark matter is a mystery in astrophysics. Some telescopes (e.g. Fermi-LAT, AMS and DAMPE) were launched to detect any possible signals of annihilating dark matter in our Galaxy. Although a certain excess of gamma rays \citep{Daylan,Calore}, anti-protons \citep{Cholis} and electrons/positrons \citep{Ambrosi,Aguilar} have been reported and claimed as dark matter signals, the uncertainties of Galactic pulsars' contributions are still too large to confirm the claims \citep{Macias}. In view of this, one particular dark matter interpretation suggests that the existence of annihilating dark matter with mass $m=48-67$ GeV annihilating via $b$ quark channel can simultaneously explain the gamma-ray and anti-proton data \citep{Daylan,Calore,Cholis}. The best-fit annihilation cross section ranges are coincident with the thermal annihilation cross section $\sigma v=2.2\times 10^{-26}$ cm$^3$ s$^{-1}$ predicted by standard cosmology \citep{Steigman}. Moreover, some later analyses of radio halos of galaxy clusters also support this suggestion \citep{Chan}. On the other hand, recent gamma-ray studies of the Omega Centauri and 47 Tuc clusters suggest a slightly smaller best-fit dark matter mass range $m \approx 30-35$ GeV and smaller annihilation cross sections with $b\bar{b}$ channel \citep{Brown,Brown2}. Therefore, combining the above suggestions, the range $m \approx 30-50$ GeV has become one of the most attentive possible range of annihilating dark matter mass. Interestingly, this narrow range of dark matter mass just satisfies the stringent limits of the Fermi-LAT gamma-ray observations of dwarf galaxies for the $b\bar{b}$ quark channel \citep{Albert,Cholis}.

In this article, we re-visit the archival radio continuum spectral data of the Abell 4038 cluster obtained by several radio observational studies \citep{Kale}. We surprisingly find a possible radio signal of annihilating dark matter manifested in the radio spectrum of the Abell 4038 cluster. Large Test Statistic values TS $>45$ for four popular annihilation channels are obtained and our results also support the recent claims of dark matter mass $m \approx 30-50$ GeV annihilating via the $b\bar{b}$ channel with the thermal annihilation cross section \citep{Daylan,Calore,Cholis,Chan}.
 
\section{Dark matter annihilation model}
Dark matter annihilation would produce a large amount of high-energy electrons and positrons. The spectra of these electrons and positrons for different annihilation channels are well-determined by numerical calculations \citep{Cirelli}. These high-energy electrons and positrons would emit synchrotron radiation in radio bands when there is a strong magnetic field. Therefore, it is possible to detect dark matter annihilation signal emitted from galaxies and galaxy clusters by radio telescopes. The radio flux emitted mainly depends on the magnetic field strength $B(r)$, number density profile of thermal electrons $n(r)$, dark matter density profile $\rho_{DM}(r)$, annihilation cross section $\sigma v$, dark matter mass $m$ and annihilation channels.

The physics of synchrotron radio emission of high-energy electrons and positrons is well-known. For low redshift galaxy clusters, the average synchrotron power at frequency $\nu$ is given by \citep{Storm}
\begin{equation}
P_{\rm syn}=\int_0^\pi d\theta \frac{(\sin \theta)^2}{2} 2\pi \sqrt{3}r_em_ec\nu_gF_{\rm syn} \left(\frac{x}{\sin \theta} \right),
\end{equation}
where $\nu_g=eB/(2\pi m_ec)$, $B$ is the magnetic field strength, $r_e$ is the classical electron radius, and the quantities $x$ and $F_{\rm syn}$ are defined as
\begin{equation}
x= \frac{2 \nu}{3 \nu_g \gamma^2} \left[1+ \left(\frac{\gamma \nu_p}{\nu} \right)^2 \right]^{3/2},
\end{equation}
where $\gamma$ is the Lorentz factor of the high-energy electrons or positrons and $\nu_p=8890[n(r)/1~{\rm cm}^{-3}]^{1/2}$ Hz is the plasma frequency, and
\begin{equation}
F_{\rm syn}(y)=y \int_y^{\infty} K_{5/3}(s)ds \approx 1.25y^{1/3}e^{-y}(648+y^2)^{1/12}.
\end{equation}

Apart from the synchrotron cooling, the high-energy electrons and positrons would cool down mainly via inverse Compton scattering of the Cosmic Microwave Background photons, Bremsstrahlung radiation and Coulomb losses. The total cooling rate (in $10^{-16}$ GeV s$^{-1}$) of a high-energy electron or positron with energy $E$ is given by \citep{Colafrancesco}
\begin{equation}
\begin{aligned}
b(E)
=&0.0254E^2B^2+0.25E^2+1.51n(r)\left[0.36+\log \left(\frac{\gamma}{n(r)} \right) \right]
\\
& +6.13n(r) \left[1+\frac{1}{75} \log \left(\frac{\gamma}{n(r)} \right) \right],
\end{aligned}
\end{equation}
where $n(r)$, $E$ and $B$ are in the units of cm$^{-3}$, GeV and $\mu$G respectively. The thermal electron number density profile in a galaxy cluster is usually modeled by \citep{Chen}
\begin{equation}
n(r)=n_0 \left(1+ \frac{r^2}{r_c^2} \right)^{-3\beta/2},
\end{equation}
where $n_0$ is the central number density, $r_c$ is the scale radius and $\beta$ is the index parameter. For the magnetic field strength, theoretical models suggest that the magnetic field strength profile in a galaxy cluster follows the thermal electron density profile \citep{Dolag,Govoni}: 
\begin{equation}
B(r)=B_0 \left[ \left(1+ \frac{r^2}{r_c^2} \right)^{-3\beta/2} \right]^{\eta},
\end{equation}
where $B_0$ is the central magnetic field strength and $\eta=0.5-1.0$ is the index modeled in simulations. Recent studies show that the central magnetic field strength $B_0$ can be written in terms of $n_0$ and the central temperature of the hot gas $T_0$: $B_0 \propto \epsilon^{-1/2}n_0^{1/2}T_0^{3/4}$, with $\epsilon=0.5-1$ \citep{Govoni,Kunz}. Therefore, the magnetic field strength profile can be determined by the parameters of a galaxy cluster. 

Generally speaking, the magnetic field strength of a galaxy cluster is high enough such that most of the high-energy electrons and positrons would cool down to non-relativistic before leaving the galaxy cluster. The cooling time scale of the high-energy electrons and positrons is much smaller than their diffusion scale so that the diffusion process is insignificant in determining the radio flux emission \citep{Storm}. Therefore, we can neglect the diffusion term in the diffusion equation and the equilibrium high-energy electron or positron number density energy spectrum is given by
\begin{equation}
\frac{dn_e}{dE}=\frac{(\sigma v)[\rho_{DM}(r)]^2}{2m^2b(E)} \int_E^m \frac{dN_{e,inj}}{dE'}dE',
\end{equation}
where $dN_{e,inj}/dE'$ is the injection energy spectrum of dark matter annihilation. The dark matter density can be obtained by assuming the hot gas in hydrostatic equilibrium: 
\begin{equation}
\rho_{DM}(r)=\frac{1}{4 \pi r^2} \frac{d}{dr} \left[-\frac{kTr}{\mu m_pG} \left(\frac{d \ln n(r)}{d \ln r}+ \frac{d \ln T}{d \ln r} \right) \right],
\end{equation}
where $\mu=0.59$ is the molecular weight and $m_p$ is the proton mass. Although recent studies show that the assumption of hydrostatic equilibrium would contribute 15-20\% systematic error in the mass profile determination \citep{Biffi}, this error is relatively small and it does not affect the final results of our analysis significantly.  

Combining the above equations, the radio flux density emitted from a galaxy cluster due to dark matter annihilation is:
\begin{equation}
S_{\rm DM}(\nu)=\frac{1}{4\pi D_L^2} \int_0^R \int_{m_e}^m 2 \frac{dn_e}{dE}P_{\rm syn} dE(4\pi r^2)dr,
\end{equation}
where $D_L$ is the distance to the galaxy cluster. The factor 2 in the above equation indicates the contributions of both high-energy electrons and positrons. Here, we assume that the dark matter distribution is spherically symmetric and the distance to the galaxy cluster is very large ($>100$ Mpc) so that it is close to a point-source emission. Furthermore, simulations show that sub-structures in galaxy clusters can enhance the annihilation rate by a factor $(1+B_{\rm sub})$ \citep{Gao,Anderhalden,Marchegiani,Sanchez}. The boost factor $B_{\rm sub}$ can be represented by a parametric form in terms of the virial mass $M_{200}$ \citep{Sanchez}:
\begin{equation}
\log B_{\rm sub}=\sum_{i=0}^{5}b_i \left[\ln \frac{M_{200}}{M_{\odot}}\right]^i,
\end{equation}
where $b_0=-0.442$, $b_1=0.0796$, $b_2=-0.0025$, $b_3=4.77\times 10^{-6}$, $b_4=4.77 \times 10^{-6}$ and $b_5=-9.69\times 10^{-8}$.

\section{Data fitting}
We use the archival radio continuum spectral data of the Abell 4038 cluster obtained by several radio observational studies for analysis \citep{Kale}. We consider the total integrated flux density emitted by the Abell 4038 cluster shown in \citet{Kale}. One special feature of the radio continuum spectral data is the non-constant spectral index of the radio flux. The total integrated radio flux is mainly synchrotron radiation and cosmic rays contribute dominantly to the emission from the radio relic and two discrete sources (A4038\_10 and A4038\_11) (see Table 1) \citep{Kale}. Several models have been proposed to account for the radio spectral shapes of cosmic-ray emission, including primary electron emission models \citep{Jaffe,Rephaeli,Rephaeli2}, secondary electron emission models \citep{Dennison}, the in-situ acceleration models \citep{Jaffe,Roland,Schlickeiser} and the adiabatic compression models \citep{Enblin}. The radio continuum spectrum of the Coma cluster has been examined by these models and some good fits can be obtained \citep{Thierbach}. The primary electron emission models can be parametrized as:
\begin{equation}
S_{\rm CR}=S_{\rm CR,0} \left(\frac{\nu}{\rm GHz} \right)^{-\alpha} \left[\frac{1+(\nu_s/\rm GHz)^{\Gamma}}{1+(\nu/\nu_s)^{\Gamma}} \right],
\end{equation}
where $\Gamma=0.5$ or 1 \citep{Thierbach}. For the secondary electron emission models, they can be written as
\begin{equation}
S_{\rm CR}=S_{\rm CR,0} \left( \frac{\nu}{\rm GHz} \right)^{-\alpha}.
\end{equation}
Only two parameters are involved for the secondary electron emission models. For the in-situ acceleration models, they can be expressed as
\begin{equation}
S_{\rm CR}=S_{\rm CR,0} \left(\frac{\nu}{\rm GHz} \right)^{-\alpha} \exp(-\nu^{1/2}/\nu_s^{1/2}).
\end{equation}
In the above three parametric forms, $S_{\rm CR,0}$, $\alpha$ and $\nu_s$ are the free parameters for fitting \citep{Thierbach}. For the adiabatic compression models, there is no analytic functional form. Nevertheless, previous studies show that the radio relic emission (the major cosmic-ray source) in the Abell 4038 cluster can be described by the adiabatic compression model using numerical calculations \citep{Weeren}. We find that the functional form in Eq.~(13) can also give very good fits for the radio relic emission in the Abell 4038 cluster. The average deviation between the functional form in Eq.~(13) and the numerical calculations using the adiabatic compression model is as small as 5\% (see Fig.~1). Although Eq.~(13) could give very good fits for the radio relic emission in the Abell 4038 cluster, we will apply the above three parametric forms to perform the analysis.

For the Abell 4038 cluster, the values of the hot gas parameters are $\beta=0.541^{+0.009}_{-0.008}$, $r_c=43 \pm 2$ kpc and $n_0=0.0174 \pm 0.0003$ cm$^{-3}$ \citep{Chen}. Applying the central temperature $T_0=3.11 \pm 0.12$ keV obtained in the Chandra observations \citep{Cavagnolo}, we can get a possible range of $B_0=6.6-9.3$ $\mu$G. We show the hot gas mass density profile, the hot gas temperature profile, the magnetic field profile and the dark matter density profile of the Abell 4038 cluster in Fig.~2. 

If there exists one more radio emission source - high-energy electrons and positrons produced from dark matter annihilation, the total radio flux would be $S_{\rm tot}=S_{\rm DM}+S_{\rm CR}$. Generally speaking, the shapes of the radio continuum spectrum for different emission sources could be different. Therefore, it is possible for us to differentiate the contributions of different emission sources and determine how the additional source from dark matter annihilation improves the fits of the radio continuum spectral data. Note that the radio emissions due to the strong cosmic-ray sources (the relic, A4038\_10 and A4038\_11) are spatially asymmetric \citep{Kale} while the dark matter contribution assumed is spherically symmetric. Here, we assume that there are two emission components in the total radio flux emission: spherically symmetric emission and the asymmetric emission. The spherically symmetric emission component mainly originates from dark matter annihilation while the asymmetric emission component (e.g. the radio relic) mainly originates from cosmic rays. Note that the three parametric forms of cosmic-ray emissions used in Eqs.~(11)-(13) do not require spatial symmetry. In the following, we will consider the total integrated flux $S_{\rm tot}$ within the whole galaxy cluster (symmetric component + asymmetric component), which is a function of $\nu$ only. Also, we will see that dark matter annihilation only contributes less than 10\% of the total radio flux. Therefore, the total resultant radio flux would be spatially asymmetric as the asymmetric cosmic-ray emission dominates the total emission. Nevertheless, it is still possible for us to constrain the spherically symmetric dark matter annihilation component using this spatially asymmetric total integrated radio flux. 

Using Eq.~(9), we can predict the radio flux contributed by dark matter annihilation $S_{\rm DM}$ as a function of radio frequencies $\nu$. Here, we follow the thermal annihilation cross section $\sigma v=2.2 \times 10^{-26}$ cm$^3$ s$^{-1}$ predicted by standard cosmology \citep{Steigman}. Therefore, only one free parameter $m$ is involved in the dark matter annihilation model. From the dark matter density profile of the Abell 4038 cluster, we get $M_{200}= 1.1 \times 10^{14}M_{\odot}$ and $B_{\rm sub}=29.5$.

For each annihilation channel and dark matter mass, we can obtain a corresponding predicted radio continuum spectrum $S_{\rm tot}=S_{\rm DM}+S_{\rm CR}$. We compare the predicted $S_{\rm tot}$ with the observed radio flux spectrum of the Abell 4038 cluster. The Likelihood $L$ between the predicted and observed radio flux spectrum can be calculated. We take the null hypothesis as the radio emission without dark matter contribution (i.e. $S_{\rm tot}=S_{\rm CR}$) and the corresponding Likelihood is given by $L_0$. We compare the Likelihood functions by the Test Statistic, which is given by 
\begin{equation}
{\rm TS}=-2 \ln \left(\frac{L_0}{L} \right).
\end{equation}

Among the three parametric forms of cosmic-ray emission, we find that only the functional form of Eq.~(13) can give the largest likelihood for the null hypothesis. As mentioned above, previous studies have shown that the cosmic-ray emission dominates the radio relic in the Abell 4038 cluster (see also Table 1) \citep{Kale}. The relic emission is closely related to the presence of a shock and it can be well described by the adiabatic compression model \citep{Weeren}, which can be well-fitted by Eq.~(13) as well (see Fig.~1). Therefore, we adopt Eq.~(13) as the model of cosmic-ray emission without dark matter annihilation as the null hypothesis for comparison. We plot the graph TS against dark matter mass $m$ for four popular annihilation channels ($e^+e^-$, $\mu^+\mu^-$, $\tau^+\tau^-$ and $b\bar{b}$) with the two extreme values of the magnetic field parameters $B_0$ and $\eta$ in Fig.~3. All of the four channels can give TS values greater than 45 (more than $6 \sigma$ statistical preference). In particular, the $\tau^+\tau^-$ channel with $m=60$ GeV ($\eta=1$ and $B_0=9.3$ $\mu$G) gives TS = 58, which corresponds to $7.6\sigma$ statistical preference. If we set $6 \sigma$ statistical preference (TS = 36) as a reference line for the best-fit ranges of $m$, we get $m=17-60$ GeV, $141-192$ GeV, $77-131$ GeV and $54-111$ GeV for $b\bar{b}$, $e^+e^-$, $\mu^+\mu^-$ and $\tau^+\tau^-$ channels respectively (see Table 2 for the corresponding parameters). We also show the best-fit radio continuum spectrum for each of the four annihilation channels (see Fig.~4). Surprisingly, the range for the $b\bar{b}$ channel significantly overlaps with the ranges suggested by many previous studies of dark matter interpretation ($m \approx 30-50$ GeV), such as the Galactic Centre gamma-ray excess \citep{Daylan,Calore}, Galactic anti-proton excess \citep{Cholis}, radio spectrum of the Ophiuchus cluster \citep{Chan} and the gamma-ray spectrum of the Omega Centauri cluster and 47 Tuc cluster \citep{Brown,Brown2}. Furthermore, the ranges of $m$ obtained for the four annihilation channels can satisfy the most stringent constraints from our Milky Way and the Milky Way dwarf spheroidal satellite galaxies \citep{Cavasonza,Albert}.

Note that the spectral fit in Fig.~1 is for the radio relic emission (the major cosmic-ray source) while the spectral fits in Fig.~4 are for the total radio flux (including radio relic, A4038\_10, A4038\_11 and dark matter contribution). Although we obtain good fits by Eq.~(13) to fit the radio relic spectrum, using Eq.~(13) alone (i.e. only cosmic-ray contribution) does not give good fits to the total radio flux spectrum. Our results show that adding a dark matter contribution $S_{\rm DM}$ can get much better fits for the total radio flux spectrum (see Table 3). Following the Akaike information criterion (AIC), the difference between the AIC values $(2p_0-2 \ln L_0)-(2p-2 \ln L)$ for the null hypothesis ($S_{\rm CR}$ only, $p_0=3$) and the dark matter hypothesis ($S_{\rm CR}+S_{\rm DM}$, $p=4$) is larger than 44, which means that the dark matter hypothesis can get much better fits. Here, $p_0$ and $p$ are the number of free parameters involved in the null hypothesis and the dark matter hypothesis respectively. 

Moreover, there are slight differences among the cosmic-ray contributions $S_{\rm CR}$ in Fig.~4 for the four annihilation channels, although we have used the same functional form in Eq.~(13). It is because the free parameters involved in that functional forms are different for the four annihilation channels. These parameters would determine the position of the exponential cutoff in the large frequency regime and the spectral index in the small frequency regime. In fact, the actual cosmic-ray contributions cannot be determined independently or theoretically. As the dark matter contributions are different among the four annihilation channels, the required cosmic-ray contributions to fit for the total radio spectrum would also be different so that there are different sets of free parameters obtained for the cosmic-ray contributions (see Table 2).

We also consider the effect of uncertainties of the hot gas parameters (within the $1\sigma$ uncertainties shown in \citet{Chen}). The hot gas parameters determine the dark matter density profile. Varying these parameters within their $1\sigma$ uncertainties can give certain maximum and minimum dark matter density profiles (i.e. maximum and minimum dark matter contributions). In Fig.~5, we show the TS values against $m$ for the maximum (dashed lines) and minimum (dotted lines) dark matter contribution scenarios. Compared with our benchmark scenario (solid lines), the variations are not very large. The constrained mass ranges are somewhat larger while the TS values are slightly smaller for the maximum and minimum dark matter contribution scenarios. Therefore, the effect of the uncertainties of the hot gas parameters are not very significant.

\section{Discussion}
In this article, we have identified a possible signal manifested in the radio continuum spectral data of the Abell 4038 cluster. Recently, there are some studies using radio spectral data to constrain or examine any possible signals of dark matter annihilation \citep{Marchegiani2,Marchegiani3}. However, the signals claimed are not very strong and the corresponding uncertainties are quite large. Therefore, our results may be able to provide a clearer radio signals of dark matter annihilation. Interestingly, the predicted range of $m$ for the $b\bar{b}$ channel completely overlaps with many of the previous claims, which further supports the existence of annihilating dark matter. 

The observational uncertainties of the radio data are very small so that we are able to identify a relatively strong signal of dark matter annihilation. Besides, the uncertainties of the involved parameters (e.g. $\beta$, $n_0$), hot gas number density profile and the temperature profile are all very small. This can help provide a precise analysis for determining the possible ranges of dark matter mass. We have shown that the effects of the uncertainties of parameters are not very significant (see Fig.~5). Including these uncertainties, the best-fit ranges of $m$ are slightly larger. Varying these parameters within their $1\sigma$ uncertainties gives slightly smaller TS values. Therefore, our benchmark values of the parameters can almost give the best-fit scenarios (largest TS values).  

However, some systematic uncertainties may be involved in our analysis. The major uncertainty is that the functional form of the $S_{\rm CR}$ used in Eq.~(13) may be oversimplified. Although this functional form is empirically good for the in-situ acceleration models \citep{Thierbach} and the radio emission of the relic in the Abell 4038 cluster assuming the adiabatic compression model, it may not fully represent a universal spectral shape for all possible emissions based on these models. For example, studies of the re-acceleration processes in galaxy clusters due to turbulent effects are usually calculated numerically \citep{Brunetti,Donnert}. The resulting spectral behaviours depend on the parameters involved and no analytic solution can be obtained \citep{Donnert}. Therefore, although the functional form in Eq.~(13) has three different free parameters, it may not be able to exhaustively and precisely reproduce all possible spectral shapes. If more precise simulations or numerical calculations are used, the TS values obtained might be significantly reduced, which would weaken our conclusions. Moreover the boost factor used in this analysis, being obtained from a numerical simulation, has a large systematic uncertainty, and there are no independent observational constraints on its value. Generally speaking, different values of the boost factor might change the best-fit range of $m$ and the TS values in this analysis. Nevertheless, our results show that using radio continuum spectral data of galaxy clusters is another excellent way to search for dark matter signals. Although the results are subject to the limitation of the abovementioned systematic uncertainties, our best-fit ranges of $m$ are consistent with the other observations and constraints. Further investigations following this direction can verify our results and help solve the dark matter mystery.

\begin{figure}
\vskip 3mm
 \includegraphics[width=80mm]{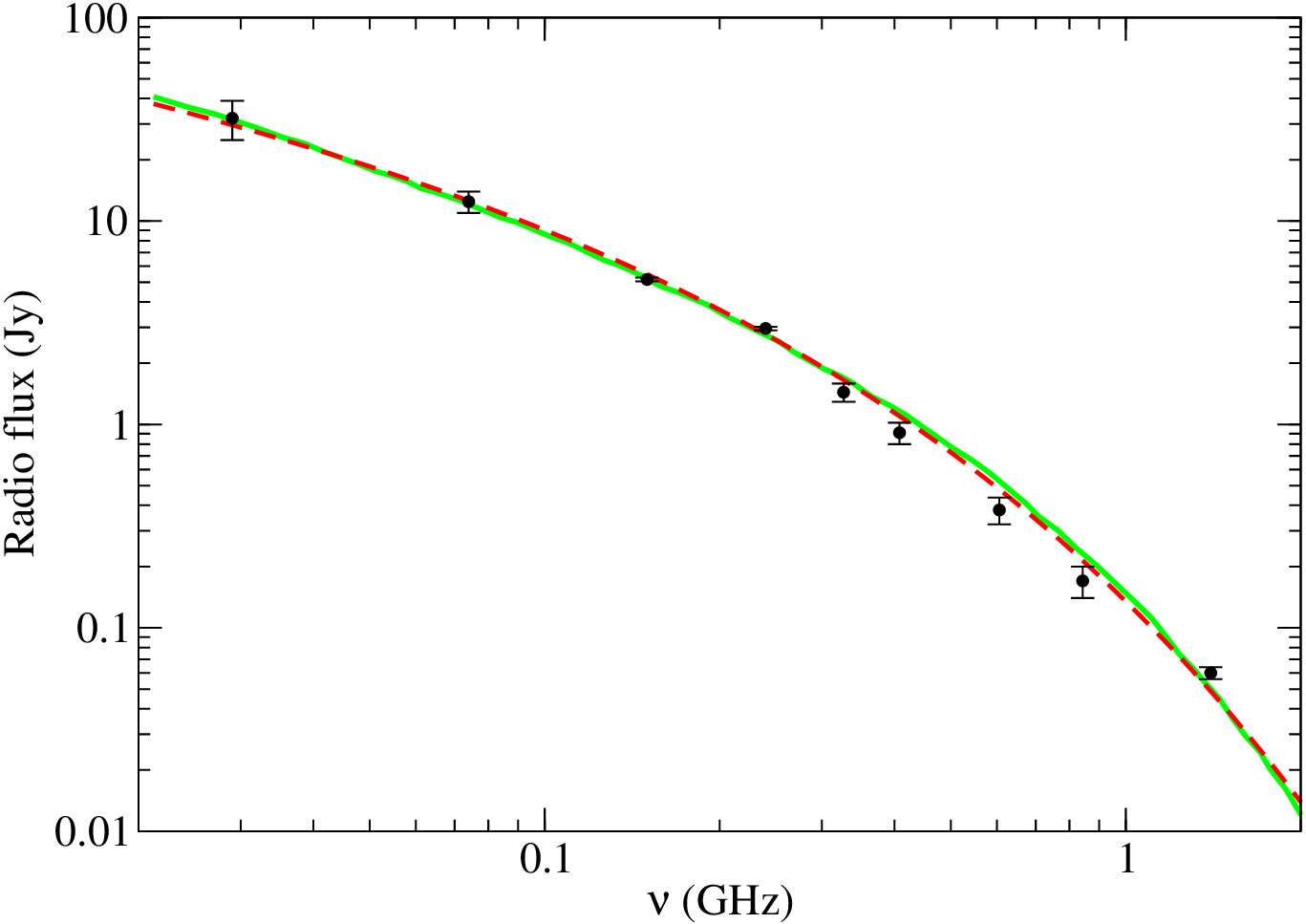}
 \caption{The radio flux spectrum of the radio relic in the Abell 4038 cluster. The data are extracted from \citet{Kale} (see Table 1). The green solid line is the best-fit spectrum obtained using the adiabatic compression model \citep{Enblin}. The red dashed line is the spectrum described by the functional form in Eq.~(13) (best-fit parameters: $S_{\rm CR,0}=17.0$ Jy, $\alpha=0.39$ and $\nu_s=0.043$ GHz).}
\vskip 3mm
\end{figure}

\begin{figure}
\vskip 3mm
 \includegraphics[width=80mm]{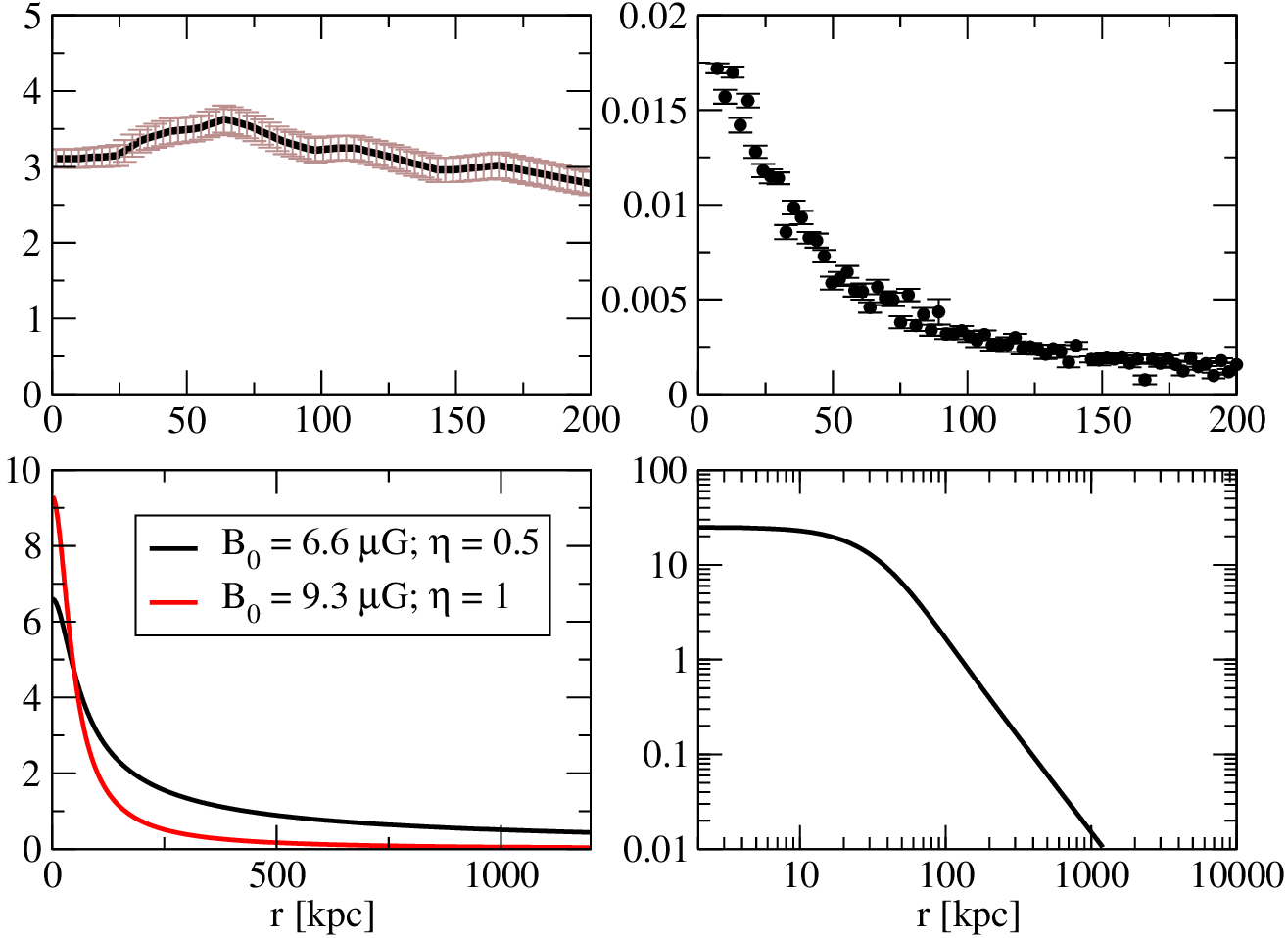}
 \caption{Different important profiles of the Abell 4038 cluster. Top left: The temperature profile of the hot gas (in keV). Top right: The number density profile of the hot gas (in cm$^{-3}$). Bottom left: The magnetic field strength profile (in $\mu$G). Bottom right: The dark matter density profile (in $10^6M_{\odot}/{\rm kpc}^3$). The data of the hot gas are extracted from the Chandra observations \citep{Cavagnolo}.}
\vskip 3mm
\end{figure}

\begin{figure}
\vskip 3mm
 \includegraphics[width=80mm]{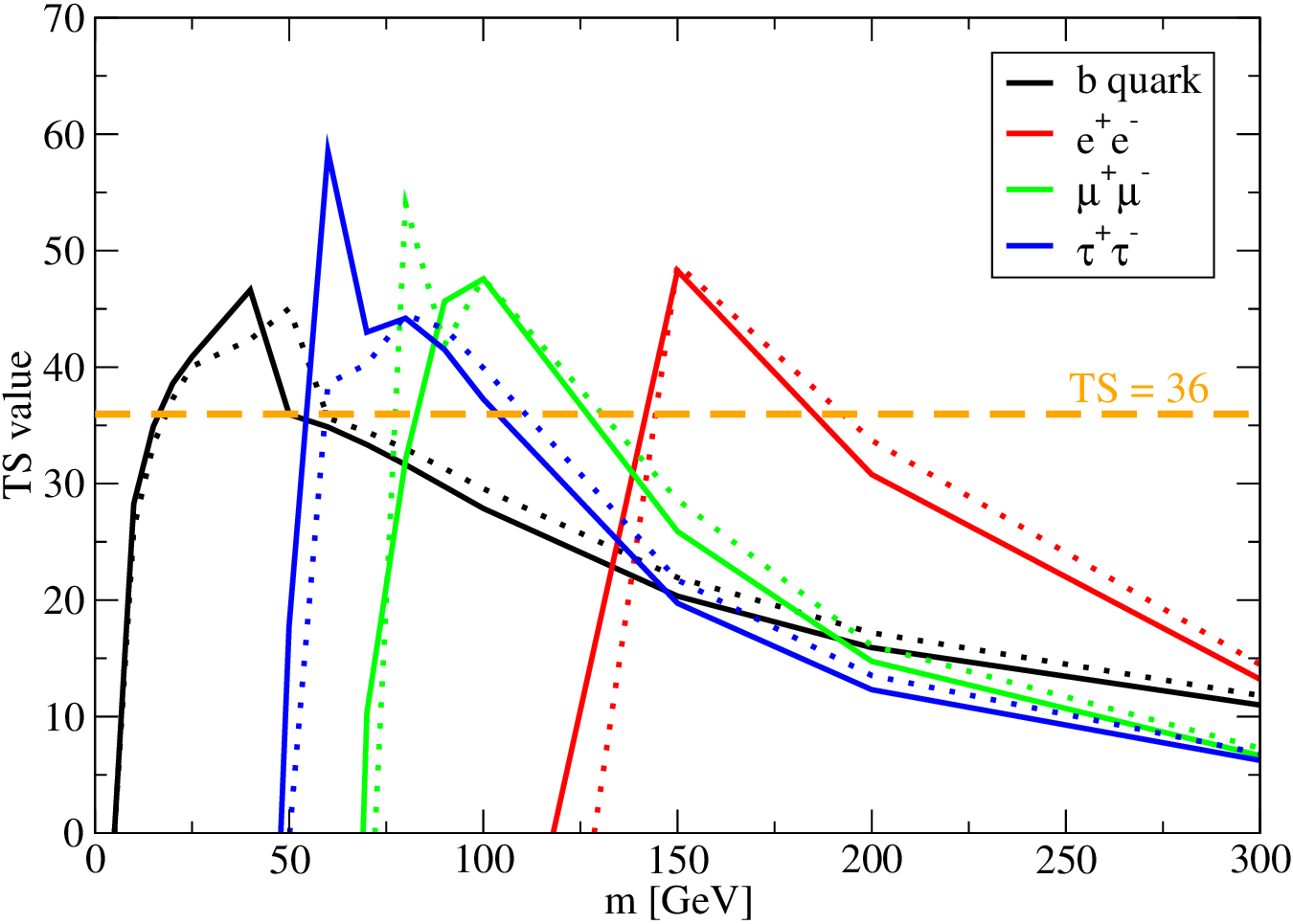}
 \caption{The TS values as a function of $m$ for four annihilation channels. The dotted lines represent the fits with $B_0=6.6$ $\mu$G and $\eta=0.5$ and the solid lines represent the fits with $B_0=9.3$ $\mu$G and $\eta=1$.}
\vskip 3mm
\end{figure}

\begin{figure}
\vskip 3mm
 \includegraphics[width=80mm]{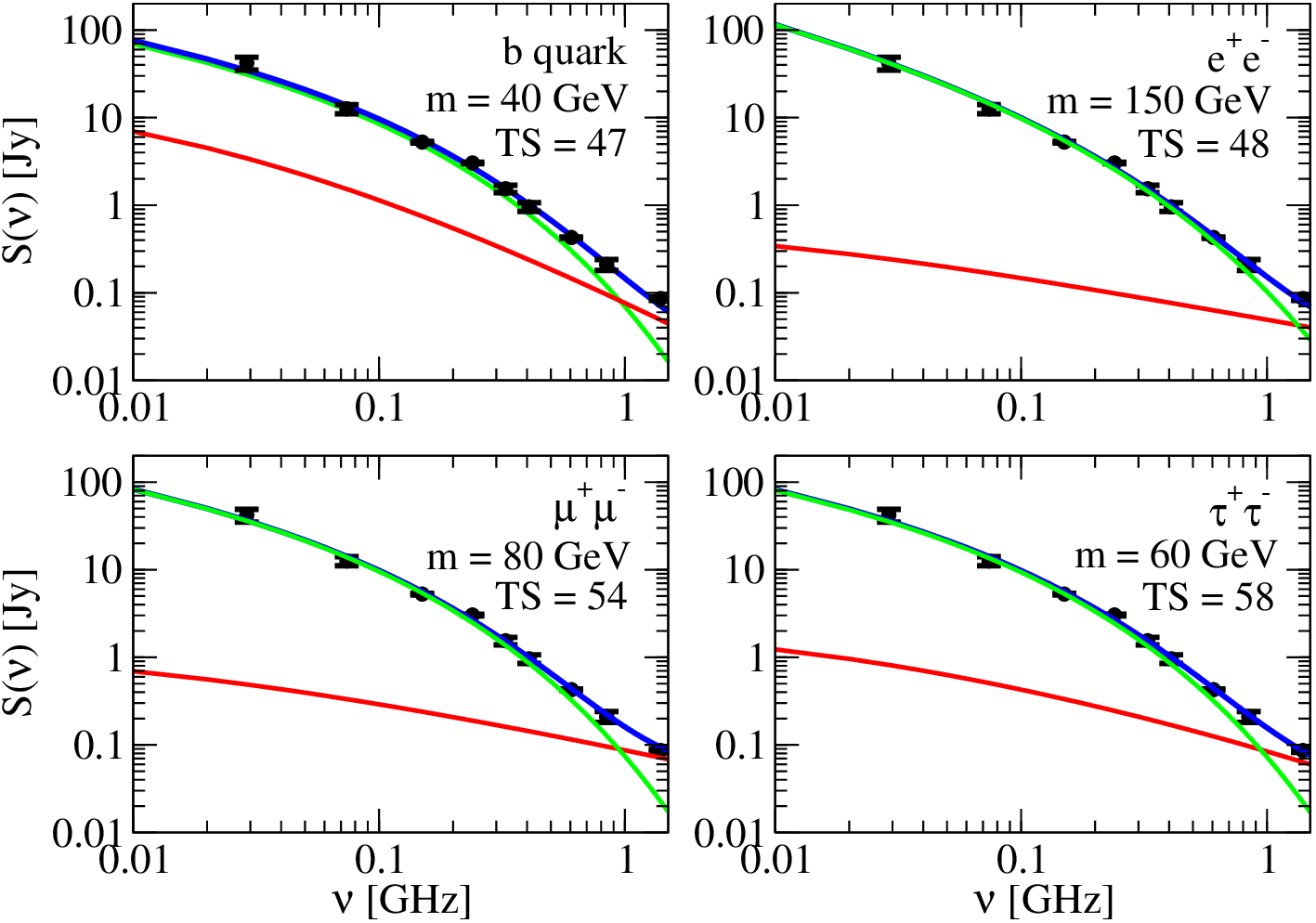}
 \caption{The best-fit spectra for four annihilation channels. The blue line is the total radio flux $S_{\rm tot}$. The green and red lines are the contributions of cosmic rays $S_{\rm CR}$ and dark matter $S_{\rm DM}$ respectively. The parameters used are shown in Table 2. The data are the total integrated radio flux of the Abell 4038 cluster \citep{Kale}.}
\vskip 3mm
\end{figure}

\begin{figure}
\vskip 3mm
 \includegraphics[width=80mm]{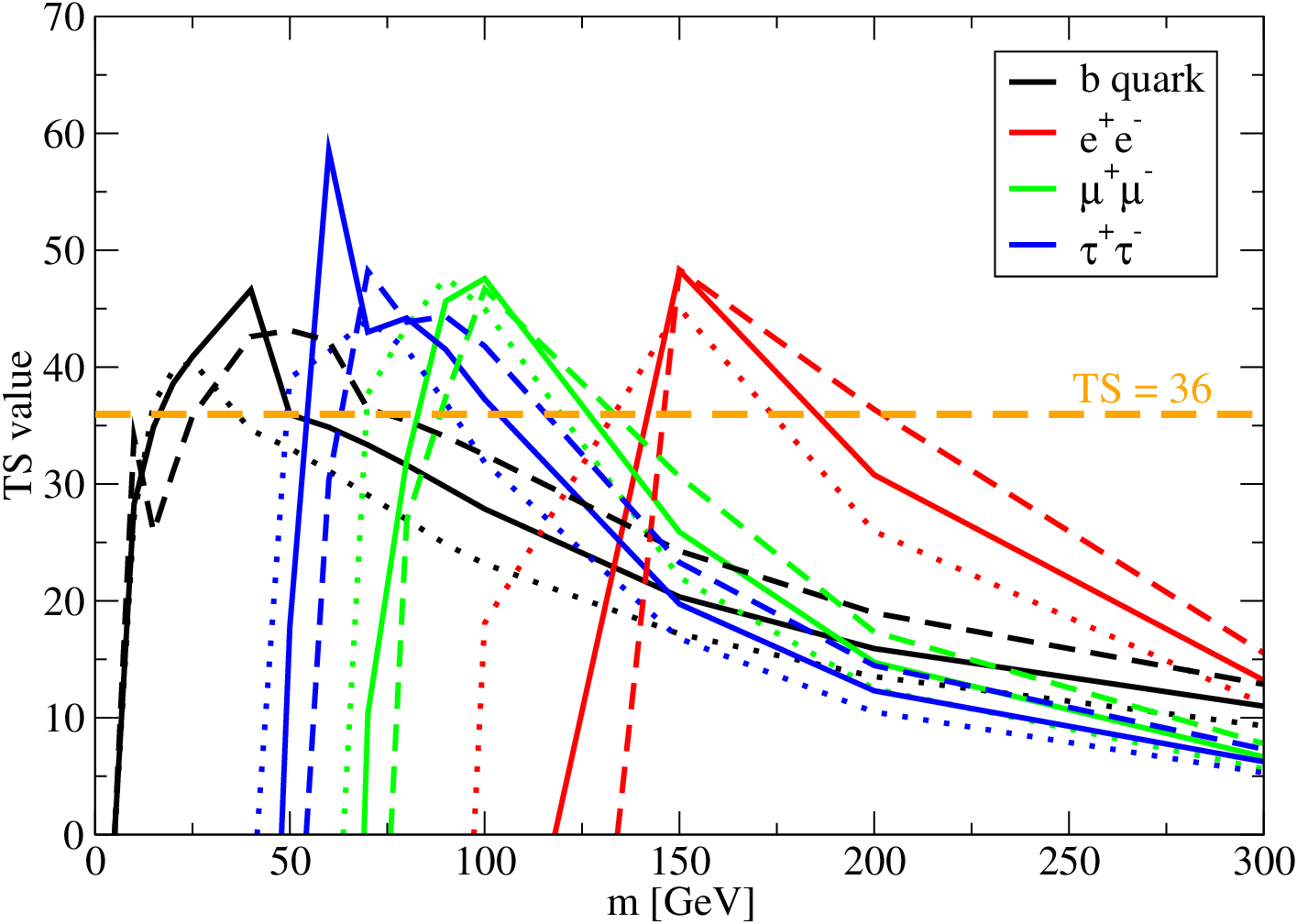}
 \caption{The TS values as a function of $m$ for four annihilation channels, including the $1\sigma$ uncertainties of the hot gas parameters. The dotted lines, solid lines and dashed lines represent the TS distributions for minimum, benchmark and maximum dark matter contributions respectively. Here, $B_0=9.3$ $\mu$G and $\eta=1$ are assumed.}
\vskip 3mm
\end{figure}

\begin{table}
\caption{The integrated radio flux density of the three major cosmic-ray sources (radio relic, A4038\_10 and A4038\_11) and the total integrated radio flux density. The data are taken from \citet{Kale}.}
 \label{table1}
 \begin{tabular}{@{}ccccc}
  \hline
  $\nu$ & Radio Relic &  A4038\_10 & A4038\_11 & Total flux \\
  (GHz) & (Jy) & (mJy) & (mJy) &  (Jy) \\
  \hline
  0.029 & $32 \pm 7$ & - & - & $42 \pm 7$ \\
  0.074 & $12.45 \pm 1.5$ & $24 \pm 5$ & $117 \pm 20$ & $12.6 \pm 1.5$ \\
  0.150 & $5.16 \pm 0.11$ & $13 \pm 2$ & $84 \pm 10$ & $5.26 \pm 0.11$ \\
  0.240 & $2.96 \pm 0.06$ & $9.0 \pm 2.0$ & $65 \pm 5$ & $3.04 \pm 0.06$ \\
  0.327 & $1.44 \pm 0.15$ & - & - & $1.54 \pm 0.15$ \\
  0.408 & $0.91 \pm 0.11$ & - & - & $0.96 \pm 0.11$ \\
  0.606 & $0.380 \pm 0.057$ & $4.5 \pm 0.8$ & $39.0 \pm 3.0$ & $0.427 \pm 0.008$ \\
  0.843 & $0.17 \pm 0.03$ & - & - & $0.21 \pm 0.03$ \\
  1.400 & $0.060 \pm 0.004$ & $2.3 \pm 0.1$ & $23.0 \pm 0.1$ & $0.086 \pm 0.004$ \\ 
  \hline
 \end{tabular}
\end{table}

\begin{table}
\caption{The best-fit parameters for the four annihilation channels. The units for $S_{\rm CR,0}$, $\nu_s$ and $B_0$ are in Jy, GHz and $\mu$G respectively.}
 \label{table2}
 \begin{tabular}{@{}lccccccc}
  \hline
  Channel & $m$ (GeV) &  TS & $S_{\rm CR,0}$ & $\alpha$ & $\nu_s$ & $B_0$ & $\eta$ \\
  \hline
  $b\bar{b}$ & 40 & 47 & 22.5 & 0.37 & 0.03 & 9.3 & 1 \\
  $e^+e^-$ & 150 & 48 & 9.02 & 0.65 & 0.05 & 9.3 & 1 \\
  $\mu^+\mu^-$ & 80 & 54 & 8.82 & 0.66 & 0.05 & 6.6 & 0.5 \\
  $\tau^+\tau^-$ & 60 & 58 & 23.4 & 0.40 & 0.03 & 9.3 & 1 \\
  \hline
 \end{tabular}
\end{table}

\begin{table}
\caption{The relative log-likelihoods (natural logarithm) for the cosmic-ray component alone model (the null hypothesis) and the best-fit cosmic-ray plus dark matter model.}
 \label{table3}
 \begin{tabular}{@{}lccc}
  \hline
  Component & Channel & $m$ (GeV) & Relative log-likelihood \\
  \hline
  \hline
  $S_{\rm CR}$ &  &  & -56.4 \\
  \hline
  $S_{\rm CR}+S_{\rm DM}$ & $b\bar{b}$ & 40 & -33.2 \\
						  & $e^+e^-$ & 150 & -32.2 \\
                          & $\mu^+\mu^-$ & 80 & -29.5 \\
                          & $\tau^+\tau^-$ & 60 & -27.5 \\
  \hline
 \end{tabular}
\end{table}

\section{acknowledgements}
The work described in this paper was supported by a grant from the Research Grants Council of the Hong Kong Special Administrative Region, China (Project No. EdUHK 28300518) and the Internal Research Fund from The Education University of Hong Kong (RG 2/2019-2020R).

\section{Data availability statement}
The data underlying this article will be shared on reasonable request to the corresponding author.

\label{lastpage}

\end{document}